\documentclass{sig-alternate-2013}

\PassOptionsToPackage{hyphens}{url}\usepackage{url}
\usepackage{comment}
\usepackage{graphicx}
\usepackage{grffile}
\usepackage{natbib}
\usepackage{verbatim}
\usepackage[caption=false]{subfig}
\usepackage{siunitx}

\newcommand{\fref}[1]{Figure~\ref{#1}}

\newcommand\pN{\mathcal{N}}
\newcommand{\xhdr}[1]{\vspace{1.7mm}\noindent{{\bf #1.}}}
\newcommand{\betadist}{Beta}
\newcommand{\adjustitemspace}{\setlength\itemsep{0em}}

\newcommand{\rot}[1]{\rlap{\rotatebox{90}{#1}}}
\newcommand{\chk}{\rlap{$\checkmark$}}

\newcommand{\domain}{\emph{Domain}}
\newcommand{\tweettime}{\emph{Tweet time}}
\newcommand{\spamscore}{\emph{Spam score}}
\newcommand{\tweetcategory}{\emph{Category}}
\newcommand{\followers}{\emph{Followers}}
\newcommand{\friends}{\emph{Friends}}
\newcommand{\statuses}{\emph{Statuses}}
\newcommand{\usertime}{\emph{User time}}
\newcommand{\tweettopic}{\emph{Tweet topic}}
\newcommand{\urlsuccess}{\emph{Past url success}}
\newcommand{\usertopic}{\emph{User topic}}
\newcommand{\usersuccess}{\emph{Past user success}}
\newcommand{\interaction}{\emph{Topic interaction}}

\permission{Copyright is held by the International World Wide Web Conference Committee (IW3C2). IW3C2 reserves the right to provide a hyperlink to the author's site if the Material is used in electronic media.}
\conferenceinfo{WWW 2016,}{April 11--15, 2016, Montr\'eal, Qu\'ebec, Canada.}
\copyrightetc{ACM \the\acmcopyr}
\crdata{978-1-4503-4143-1/16/04. \\
http://dx.doi.org/10.1145/2872427.2883001}
\begin{document}
\pagenumbering{arabic}

\title{Exploring limits to prediction in complex social systems} 
\subtitle{}
\numberofauthors{5} 
%
\author{
%
%
\alignauthor
Travis Martin\\
       \affaddr{University of Michigan}\\
       \affaddr{Dept. of Computer Science} \\
       \affaddr{Ann Arbor, MI} \\
       \email{travisbm@umich.edu}
\alignauthor
Jake M. Hofman\\
       \affaddr{Microsoft Research}\\
       \affaddr{641 6th Ave, Floor 7}\\
       \affaddr{New York, NY}\\
       \email{jmh@microsoft.com}
\and \alignauthor
Amit Sharma\\
       \affaddr{Microsoft Research}\\
       \email{amshar@microsoft.com}\\
   \alignauthor
Ashton Anderson\\
       \affaddr{Microsoft Research}\\
       \email{ashton@microsoft.com}
\alignauthor
Duncan J. Watts \\
       \affaddr{Microsoft Research}\\
       \email{duncan@microsoft.com}
}

\maketitle

\begin{abstract}
How predictable is success in complex social systems?
In spite of a recent profusion of prediction studies that exploit online social and information network data, this question remains unanswered, in part because it has not been adequately specified.
In this paper we attempt to clarify the question by presenting a simple stylized model of success that attributes prediction error to one of two generic sources: 
insufficiency of available data and/or models on the one hand; and inherent unpredictability of complex social systems on the other.
We then use this model to motivate an illustrative empirical study of information cascade size prediction on Twitter.
Despite an unprecedented volume of information about users, content, and past performance, our best performing models can explain less than half of the variance in cascade sizes. 
In turn, this result suggests that even with unlimited data predictive performance would be bounded well below deterministic accuracy.
Finally, we explore this potential bound theoretically using simulations of a diffusion process on a random scale free network similar to Twitter.
We show that although higher predictive power is possible in theory, such performance requires a homogeneous system and perfect \emph{ex-ante} knowledge of it: even a small degree of uncertainty in estimating product quality or slight variation in quality across products leads to substantially more restrictive bounds on predictability.
We conclude that realistic bounds on predictive accuracy are not dissimilar from those we have obtained empirically, and that such bounds for other complex social systems for which data is more difficult to obtain are likely even lower.
\end{abstract}




\section{Introduction}
From the motions of the planets to the vagaries of the weather to the movements of the economy and financial markets to the outcomes of elections, sporting events, and Hollywood awards nights, prediction is of longstanding interest to scientists, policy makers, and the general public~\cite{sherden1998,orrell2008,tetlock2015}. 
The science of prediction has made enormous progress in domains of physical and engineering science for which the behavior of the corresponding systems can be well approximated by relatively simple, deterministic equations of motion~\cite{weaver1961}. 
More recently, impressive gains have been made in predicting short-term weather patterns~\cite{bauer2015}, demonstrating that under some conditions and with appropriate effort useful predictions can be obtained even for extremely complex and stochastic systems.

In light of this history it is only natural to suspect that social and economic phenomena can also be brought within the sphere of scientific prediction. Although such a development has been anticipated since at least the days of Newton~\cite{watts2011,tetlock2015}, the track record of social and economic predictions has been marred by competing and often contradictory claims. 
On the one hand, proponents of various methods have claimed to accurately predict a wide variety of phenomena, ranging from the box-office success of movies~\cite{asur2010}, to the outcomes of political conflicts~\cite{demesquita2010} and social trends~\cite{friedman2010}, to the spread of epidemics~\cite{hufnagel2004,colizza2007} and ``viral'' products~\cite{berger2013}, to the stock market~\cite{bollen2011}.
On the other hand, critics have contended that claims of success often paper over track records of failure~\cite{sherden1998}, that
expert predictions are no better than random
\cite{tetlock2005, gardner2010},
that most predictions are wrong
\cite{schnaars1989,devany2004,parish2006},
and even that predicting social and economic phenomena of any importance is essentially impossible \cite{taleb2010}.
Repeated attempts to deflate expectations notwithstanding, the steady arrival of new methods---game theory~\cite{demesquita2010}, prediction markets~\cite{surowiecki2005,arrow2008}, and machine learning~\cite{domingos2015}---along with new sources of data---search logs~\cite{choi2012}, social media~\cite{asur2010,bollen2011}, MRI scans~\cite{berns2012}---inevitably restore hope that accurate predictions are just around the corner.

\xhdr{Characterizing predictability} Adjudicating these competing claims is difficult, in part because they are often stated in vague and inconsistent terms, and hence are impossible to evaluate either individually or in comparison with one another. 
For example, predicting the box office revenue for a feature film or the number of flu cases in a given city days in advance is a very different matter than predicting the next blockbuster movie or avian flu pandemic months in advance. 
It is therefore clearly misleading to cite performance on ``easy'' cases as evidence that more challenging outcomes are equally predictable; yet precisely such conflation is practiced routinely by advocates of various methods, albeit often implicitly through the use of rhetorical flourishes and other imprecise language. On the other hand, it is also misleading to imply that even if extreme events such as financial crises and societal revolutions cannot be predicted with any useful accuracy~\cite{taleb2010}, predictive modeling is counterproductive in general.  

Compounding the lack of clarity in the claims themselves is an absence of a consistent and rigorous evaluation framework. For example, it is well understood in theory that predictive accuracy cannot be reliably estimated from isolated predictions, especially when selected \emph{ex-post}~\cite{tetlock2005}. Likewise, is uncontroversial to state that model performance can be evaluated only with respect to the most relevant baseline~\cite{goel2010a}, or that incremental performance improvements do not necessarily translate to meaningful improvements in the outcome of interest~\cite{goel2010b}. In practice, however, claims of predictive accuracy (or inaccuracy) are rarely subject to such scrutiny, hence many claims that violate one or more of these conditions are allowed to stand uncontested.

Together, inadequate problem specification and inconsistent evaluation criteria have obscured a question of fundamental importance: To the extent that predictions are less accurate than desired, is it simply that the existing combination of methods and data is insufficient; or is it that the phenomenon itself is to some extent inherently unpredictable~\cite{salganik2006}? 
Although both explanations may produce the same result in a given context, their implications are qualitatively different---the former implies that with sufficient ingenuity and/or effort, failures of prediction can in principle always be corrected, whereas the latter implies a performance limit beyond which even theoretically perfect predictions cannot progress~\cite{watts2011}. In other words, if socioeconomic predictions are more like predicting a die roll than the return time of a comet, then even a ``perfect'' prediction would yield only an expected probability of success, leaving a potentially large residual error with respect to individual outcomes that could not be reduced with any amount of additional information or improved modeling~\cite{salganik2006,holme2015}.  

\xhdr{Ex-ante prediction} In this paper we investigate this question in a specific sub-domain of socioeconomic phenomena, namely predicting success. 
More specifically, we focus on predicting success \emph{ex-ante}, meaning that we are concerned with the accuracy of predictions that are made prior to the events of interest themselves.
This restriction may sound tautological, but it highlights an important source of ambiguity in the recent literature about prediction, much of which has focused on predicting outcomes that unfold over time, such as retweet cascades on Twitter, likes on Facebook, or views on YouTube. 
In such domains it has become increasingly popular to adopt what has been recently labeled a ``peeking strategy''~\cite{shulman2016}, in which one predicts some property of a dynamic process after having observed the state of the same process at some earlier stage 
\cite{jamali2009,szabo2010,lerman2010,romero2013,weng2014,cheng2014,pinto2013,maity2015,zhao2015,yu2015}. 
As has been shown elsewhere~\cite{shulman2016}, peeking strategies in general 
perform much better than strategies that rely exclusively on \emph{ex-ante} observable features alone. Moreover, as has also been argued elsewhere~\cite{watts2011,cheng2014}, peeking strategies can be useful in practice, say by allowing marketers to adjust their promotional efforts mid-stream in response to early indications of success or failure. 

Nevertheless, predictions based on peeking differ fundamentally from \emph{ex-ante} predictions, which by definition rely exclusively on features---whether of the object itself, the environment, or some combination of the two---that could have been known, and hence manipulated, prior to the process itself commencing.
By contrast, peeking strategies derive their power from cumulative advantage dynamics~\cite{shulman2016}, according to which entities that are successful early on tend to be successful later on as well, regardless of any intrinsically differentiating attributes such as higher quality or contextual appeal~\cite{watts2011}. 
The difference between these two approaches can be clarified by considering how a prediction can be interpreted: whereas \emph{ex-ante} predictions claim, in effect, that ``X will succeed because it has properties A, B, and C'', peeking strategies instead claim that ``X will succeed tomorrow because it is successful today.'' Although both types of predictions can be informative, only \emph{ex-ante} predictions offer actionable guidance on how to optimize for success during the creation process.
Further, it is precisely this guidance that motivates much of the interest in prediction---namely the potential to create successful content, products, ideas, etc., by manipulating features that are predictive of success.

To illustrate, consider the exercise of predicting total box office revenues for a movie after having observed its opening weekend, a feature that has long been known to be highly informative~\cite{simonoff2000}. Although arguably still useful---say for modifying marketing and distribution plans---conditioning one's prediction on opening weekend does not address how to make and market a successful film in the first place. 
As appealing as peeking strategies may be from a performance perspective, it is our contention that when people talk about prediction they are generally referring to \emph{ex-ante} prediction; therefore, it is \emph{ex-ante} prediction on which we focus here.

\xhdr{Our contributions} This paper makes three contributions.
First, after reviewing related work, we articulate a stylized model of success that distinguishes between the two sources of \emph{ex-ante} predictive failure outlined above: errors in the predictive model versus intrinsic unpredictability in the system.
The model formalizes intuitive notions of ``skill'' and ``luck'' and their relative contributions to success, and suggests a natural metric for evaluating performance when predicting success.

Second, we use this model to motivate an empirical study of predicting the size of retweet cascades on Twitter.
Although Twitter cascades are ultimately just one of many specific domains in which prediction is practiced, it is a topic that has attracted considerable recent interest in the computational social science literature~\cite{bakshy2011,cheng2014}. It is also a domain for which we have an exceptionally large amount of data, both in terms of number of observations and number of features; thus if accurate prediction is possible in any complex social system it ought to be possible here. As a consequence, we believe that the particular empirical case that we study here serves as a conceptually clear illustration of the more general theoretical point we wish to make.
Considering a series of increasingly powerful models, we find that the best such model exhibits better out-of-sample performance than has previously been reported~\cite{bakshy2011}. We note, however, that the model's success relies heavily on an exceptionally informative feature set, including the past performance of identical pieces of content, that is rarely available outside of social media platforms like Twitter. Moreover, even under these generous conditions, our model is able to explain less than half of the variance in cascade sizes, suggesting that predictions of success in general are far from deterministic.

Third, we conduct a simulation study of a spreading process on a random scale-free network similar to that used previously to model cascades on Twitter~\cite{goel2015}. We show that for some parameter values and under the assumption of perfect \emph{ex-ante} knowledge about the state of the system, it is possible to achieve much higher performance than any model has yet realized. We also show, however, that predictive performance is highly sensitive even to small errors in \emph{ex-ante} knowledge as well as to increased heterogeneity in product quality. For both reasons we conclude that practical limits to prediction are likely closer to those attained in our empirical study than the bounds suggested by our idealized simulations. 
 
%


\section{Related Work}
\label{sec:related}
The recent proliferation of online data---in particular deriving from search activity and user-actions on social media sites---has driven a flurry of efforts to mine such data to predict a range of online and also offline behavior.
With respect to offline behavior, Polgreen et al.~\citep{polgreen2008} and Ginsberg et al.~\cite{ginsberg2009} used search logs to ``predict'' U.S. Center of Disease Control (CDC) caseload reports two weeks in advance of their publication, while Goel et al.~\cite{goel2010a} used a similar approach to predict movie box office revenues, song rankings, and video sales one week to one month in advance. 
Contemporaneously, Asur et al.~\cite{asur2010} used counts of Twitter mentions to predict movie box office revenues, while Bollen et al.~\citep{bollen2011}, also studying Twitter posts, claimed that user sentiment could predict price fluctuations in related stocks---a claim that subsequently led to the creation of a ``Twitter-based hedge fund.'' 
Although some of the more dramatic claims arising from this early work have subsequently been criticized~\cite{goel2010a,lazer2014} the general idea of correlating online activity to offline outcomes continues to attract interest~\cite{choi2012}.

With respect to online behavior, Bakshy et al.~\cite{bakshy2011} used a combination of content and user features to predict the size of Twitter cascades in which tweets that included URLs propagated via retweets along the follower graph. Bakshy et al.\ reached three main conclusions that are germane to the current discussion: first, that by far the most informative feature was the past success of the seed user; second, that after factoring in past success neither additional seed features nor content features added appreciably to model performance; and third, that even the best performing model was able to explain only about a third of the observed variance $(R^2 \approx 0.34)$. This last result is notable because the model itself was extremely well calibrated---i.e.\ \emph{on average} predicted and actual cascade size were highly correlated $(R^2 \approx 0.98)$---implying that errors in predicting individual outcomes derive at least in part from intrinsic stochasticity in the underlying generative process~\cite{salganik2006,holme2015}, and hence are to some extent ineradicable.

Other authors, meanwhile, have reached more optimistic conclusions. For example, Hong and Davidson~\cite{hong2010} introduced a classifier based on content features that predicted whether a given user would be retweeted in the future, claiming significant improvements over various baselines. Petrovic et al.~\cite{petrovic2011} introduced a different classifier to predict whether or not a given tweet would be retweeted, in this case claiming a ``huge'' improvement over relevant baselines. Jenders et al.~\cite{jenders2013} introduced a learning model that they claimed predicted ``viral tweets with high accuracy.'' Szabo and Hubmerman~\cite{szabo2010} presented a method for ``accurately predicting the long time popularity of online content'', Weng~\cite{weng2013} examined the role of network structure in forecasting cascades, claiming that ``future popularity of a meme can be predicted by quantifying its early spreading pattern in terms of community concentration,'' and Cheng et al.~\cite{cheng2014} claimed ``strong performance in predicting whether a cascade will continue to grow in the future.''

As with the broader prediction literature surveyed above, however, the combination of imprecise language along with differences in modeling approaches, prediction targets, and performance metrics greatly complicates the exercise of evaluating or comparing competing claims.
For example, although Hong and Davidson's model does arguably outperform various baselines, all the methods they study exhibit precision and recall less than 50\%; thus an  alternative interpretation of their results would be that content features are largely uninformative---a result that would be completely consistent with Bakshy et al. Likewise, Petrovic et al.\ beat a random classifier but their model's $F1$ score remains below 50\%.
Jenders et al.'s ``high accuracy,'' meanwhile, refers only to predicting tweets that will achieve more than some threshold $T$ number of retweets---a much easier task than predicting cascade size over the whole data set, and one on which both their model and also a na{\"i}ve baseline model do extremely well. And both Szabo and Huberman and Cheng et al.\ along with several others~\cite{lerman2010,jamali2009,romero2013,weng2014,cheng2014,pinto2013,maity2015,zhao2015,yu2015} utilize different versions of the peeking strategy discussed above.

An important consequence of this lack of consistency across studies, with respect to the data sets considered, the quantities about which to make predictions, and the performance metrics used to evaluate these predictions, is that it is essentially impossible to assess whether predictive accuracy is improving in a meaningful way or is approaching any limits\footnote{The problem of changing prediction criteria has been discussed separately in the learning literature under the rubric of overfitting. See, for example, \url{http://hunch.net/?p=22}}.
We conclude that in spite of the recent attention devoted to prediction of online phenomena like cascades and offline phenomena like consumer behavior, the central question of this paper---namely to what extent errors in prediction represent inadequate models and/or data versus intrinsic uncertainty in the underlying generative process---remains unanswered.


\section{A Stylized Model of Success}
\label{sec:model}
\begin{figure}
\centering
\includegraphics[scale=0.4]{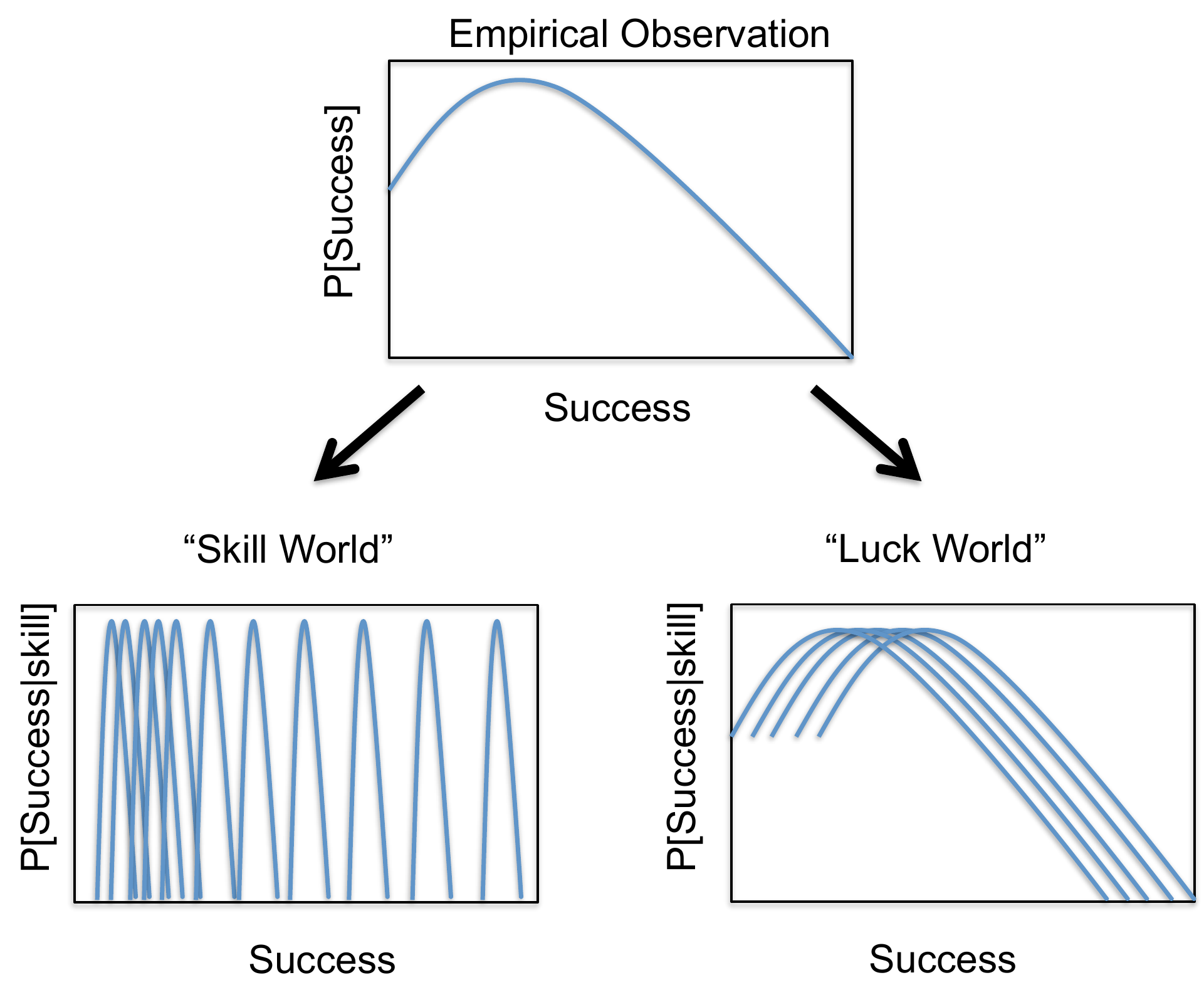}
\caption{Schematic model illustrating two stylized explanations for an empirically observed distribution of success.}
\label{fig:toy-model}
\end{figure}

To clarify the question and motivate our approach, we introduce a simple conceptual model illustrated schematically in Fig.~\ref{fig:toy-model}.
The top panel of Fig.~\ref{fig:toy-model} represents an empirically observed distribution of success. As is typical across many domains, including cultural markets~\cite{salganik2006}, scientific citations~\cite{price1965}, business firms~\cite{ijiri1977}, and personal wealth~\cite{frank2010},  the distribution is right skewed with a heavy tail, implying that most individuals/entities are relatively unsuccessful while a small minority enjoy outsize success.
Following Mauboussin~\cite{mauboussin2012}, we claim that this empirical observation can be accounted for by some mix of ``skill'' and ``luck,'' where here skill refers to some combination of stable, intrinsic attributes of the individual or entity in question and luck refers to systemic randomness that is independent of these attributes.
Our notion of skill is therefore extremely broad, covering related concepts such as quality, appeal, and potential. It can also depend on time, context, or other features of the environment, and in general will not be directly observable---although as we will argue later, one would expect that many of its contributing features will be.
Correspondingly, our notion of luck is also extremely broad, encompassing essentially all forms of intrinsic stochasticity.

Consistent with Mauboussin~\cite{mauboussin2012}, we note that the relative contribution of luck versus skill to overall success can vary between two extremes: in one extreme, pictured in the lower left panel of Fig.~\ref{fig:toy-model}, success is accounted for almost entirely by skill; whereas in the other extreme (Fig.~\ref{fig:toy-model}, lower right panel) it is accounted for almost entirely by luck.
In the ``skill world,'' that is, the variance of success \emph{once conditioned on skill} is extremely small, hence almost all the variance in the overall distribution can be attributed to differences in skill. In the ``luck world,'' by contrast, success conditioned on skill exhibits almost the same variance as the original distribution, hence differences in skill contribute relatively little to overall success.
According to Mauboussin, individual sports such as tennis and track and field are clustered at the skill end of the continuum, roulette and other forms of gambling are clustered at the luck end, and most real-world activities, including team sports, investing, and business activities fall somewhere in-between.

Mauboussin's luck-skill tradeoff can be formalized
by expressing the degree of skill inherent in a particular domain as the reduction in overall variance in outcomes that be can be achieved by conditioning on skill.\footnote{We note that our formulation of the skill-luck tradeoff is somewhat different than Mauboussin's, who instead focuses on reversion to the mean in performance over time. Although conceptually similar, our formulation lends itself more naturally to our current focus on prediction.}
In turn, the best possible performance for a predictive model is determined by this reduction in variance.
To clarify, imagine a hypothetical model of success $s = f(q) + \epsilon $ in which skill $(q)$ is the sole predictor and luck $(\epsilon)$ appears as uncorrelated, zero mean noise.
Assuming that skill has been correctly identified and precisely estimated,
let $F$ be the fraction of variance remaining after conditioning on skill:
\begin{equation}
F = \frac{\mathbb{E}[\textrm{Var}(\textrm{S}|Q)]}{\textrm{Var}(\textrm{S})}
  = \frac{\sum_q \sum_{i:q_i=q}(\bar{s}_q-s_i)^2}{\sum_j (\bar{s}-s_j)^2},
\label{eq:F}
\end{equation}
where $\bar{s}_q$ denotes the average success at skill level $q$ and $\bar{s}$ is the mean success over all observations.
In a pure skill world the variance in success given skill is typically much smaller than the overall variance,
hence $F \rightarrow 0$, whereas in a pure luck world these variances are approximately equal,
hence $F \rightarrow 1$.
The connection with predictive performance then follows naturally by using $\bar{s}_q = f(q)$ to rewrite $F$ as:
\begin{equation}
F = \frac{\sum_i (f(q_i)-s_i)^2}{\sum_j (\bar{s}-s_j)^2} = 1-R^2,
\label{eq:r-squared}
\end{equation}
where $R^2$ is the coefficient of determination, a standard and widely-used metric of predictive performance, often interpreted as the fraction of variance ``explained'' by the model. From this equivalence it becomes clear that our hypothetical model can attain theoretically perfect prediction in a pure skill world $(R^2 = 1)$ whereas in a pure luck world it offers no predictive value at all $(R^2 = 0)$.\footnote{Predictability can also be expressed in terms of information theory~\cite{delsole2004}. In a perfect luck world the mutual information between outcome and skill is zero, whereas in a perfect skill world the mutual information is maximal.} Depending on the importance of intrinsic randomness in determining outcomes, therefore, the theoretical limit of predictive performance can be expressed in terms of the $R^2$ for our idealized model.

Presumably no real model can satisfy the assumptions of this idealized model. As noted above, for example, ``skill'' will typically be unobservable in practice---at least directly---and the functional form of the mapping $f(q)$ to success will generally be unknown; thus the assumptions made above that predictions can be reduced to one ``master'' predictor and that this predictor can be estimated with zero error are unlikely to hold in most settings.
As we will illustrate in the next section, attempts at prediction must in practice try to approximate both skill (e.g.\ as different combinations of observable features), and $f(q)$ (e.g.\ by comparing the performance of different models such as linear regression, regression trees, random forests, etc.).
As we will show, it is possible to realize large gains in performance by including increasingly comprehensive feature sets and using increasingly sophisticated models; however it is almost certainly the case that even the best such model will achieve an $R^2$ less than the theoretical limit. Nevertheless, we argue that the notion of a limit \emph{even for a perfect model} is a useful construct against which empirical values of $R^2$ can be compared, hence for the remainder of this paper we use $R^2$ as our metric of model performance.\footnote{In addition to its natural interpretation as the fraction of explained variance, 
$R^2$ has some other desirable features as a benchmark for evaluating performance. First, unlike other natural measures such as MAE and RMSE, $R^2$ is unitless, hence it allows for meaningful comparisons across data sets and domains. If, say, the empirical distribution of cascade sizes on Twitter were to change over the course of several years---as is likely---performance measured in terms of MAE might well change even for the same model. By contrast, performance measured in terms of $R^2$ counts only the reduction in variance relative to the overall distribution, and hence is far more robust to distributional changes. Moreover, $R^2$ can be used to compare model performance across domains as different as Twitter cascades, life course outcomes, and heritability of phenotypic traits, whereas performance measured in terms of MAE/RMSE would yield wildly different numbers that would be hard to compare.
Second, unlike other unitless measures such as classification performance (e.g. precision, recall, F1 score), $R^2$ does not depend on partitioning outcomes into ad hoc and necessarily arbitrary classes (e.g. ``successful'' or ``viral'' versus ``unsuccessful''), hence is less susceptible to manipulation (e.g. by setting the threshold to make the prediction task easier).}



\section{Predicting Cascades on Twitter}
\label{sec:data}
With this stylized model of success in mind, we now turn to the specific task of forecasting the size of information cascades on Twitter.
To reiterate the argument above: to the extent that the success of an information cascade is based on ``skill'', this problem reduces to one of estimating the intrinsic quality and appeal of content and users, as well as the interaction between the two; if, however, success is driven largely by chance then even reliable estimates of skill would provide relatively little information about the final success of cascades.
As noted above, ``skill'' in this context (as in many others) is inherently unobservable and the function $f(q)$ mapping skill to success is unknown. Instead, therefore, we adopt an empirical and exploratory approach, fitting a series of increasingly complex models that capture ever-more comprehensive sets of content and user features, which we conjecture correspond to increasingly precise estimates of skill. We then evaluate the performance of these features on out-of-sample predictions of cascade size using a variety of different statistical models---specifically, linear regression, regression trees, and random forests---that we view as coarse substitutes for different functional forms of $f(q)$. 
The remainder of this section proceeds as follows: we first describe the dataset of tweets; then we describe our methods for feature extraction, along with the resulting features; and finally we report the performance of our various models.

\subsection{Data}
Recalling that our goal is to predict \textit{ex-ante} the total number of retweets for a particular cascade, given the content of the URL and properties of the user who started it, we construct a dataset of \emph{all} tweets from the month of February 2015 containing URLs.
To prevent right-censoring issues, we track retweets for another month beyond this period (until the end of March).
We resolve shortened URLs (e.g., \url{bit.ly} and \url{t.co} links) to their source address and treat multiple URLs within any given tweet as independent cascades.

We perform the following filtering steps to ensure that our dataset contains links to actual content such as news, images, or stories, not tweets produced by automated bots or spammers.  First, we use a proprietary social media classifier at Microsoft to score each tweet on its spam content and discard any tweet with a sufficiently high spam score.
This leaves us with 1.47 billion
base tweets, excluding retweets and quoted tweets.
Second, we consider only URLs that come from popular English language domains, to ensure that our dataset contains legitimate content for which we can reliably construct meaningful content features.
We consider popular domains to be those with many retweets from many unique tweeters, over a wide period of time. To construct this list, we identify the top 400 domains by retweets and unique tweeters for two months separately (February 2015 and October 2014). Including only domains that score in the top 400 for both months, we obtain 129 popular domains. Eliminating non-english language domains manually leaves 100 remaining domains which include all major providers of online content, including news (e.g., \url{nytimes.com}, \url{theguardian.com}), entertainment (e.g., \url{buzzfeed.com}, \url{onion.com}), videos (e.g., \url{youtube.com}, \url{vine.co}), images (e.g., \url{instagram.com}, \url{twitpic.com}, \url{imgur.com}) and consumer products (e.g., \url{amazon.com}), as well as generic content distributors such as \url{facebook.com} and \url{twitter.com}.
After filtering URLs by domain, the dataset contains 852M tweets by 51M users.

\begin{figure*}[t]
\center
\subfloat{\includegraphics[scale=0.5]{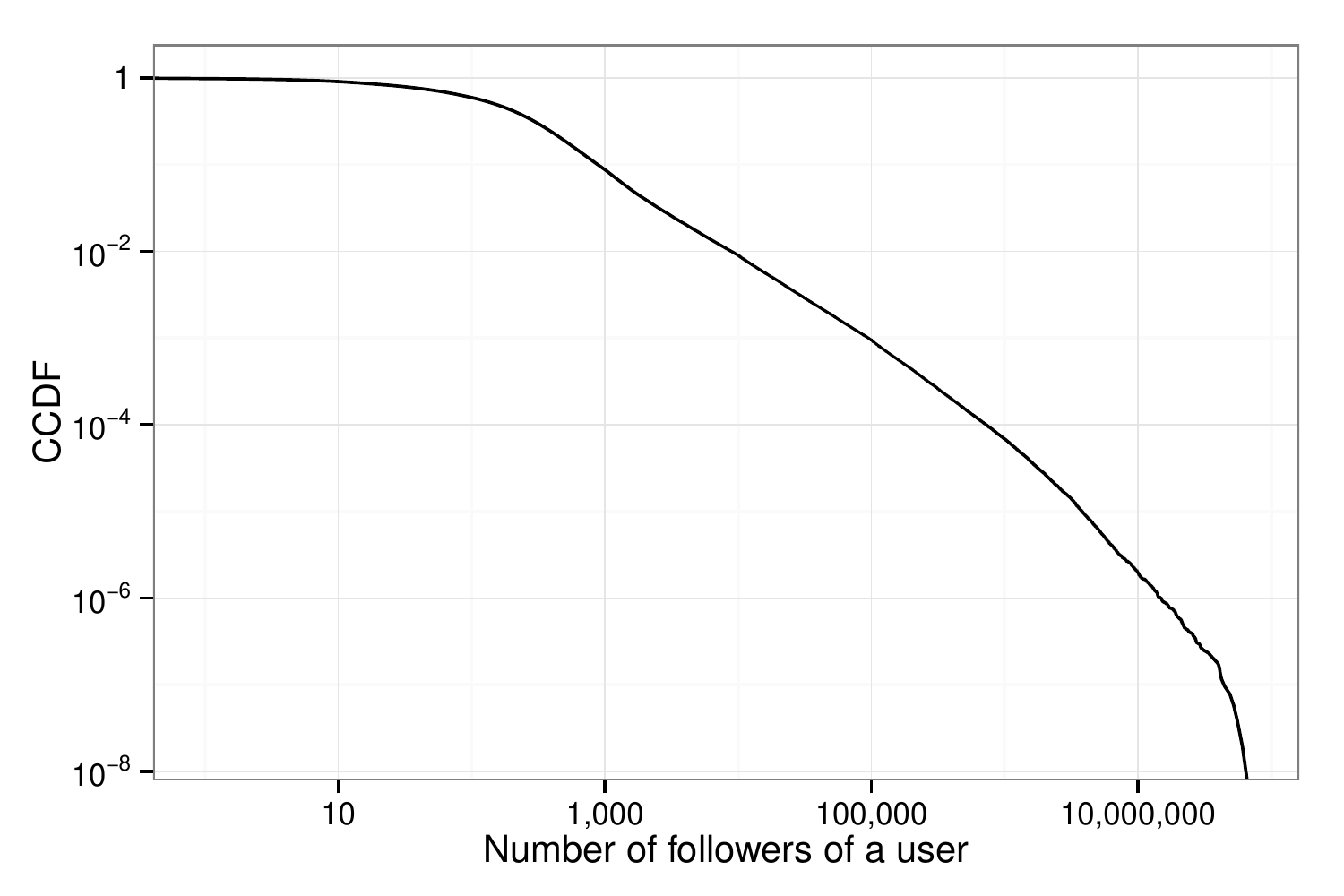}}
\quad
\subfloat{\includegraphics[scale=0.5]{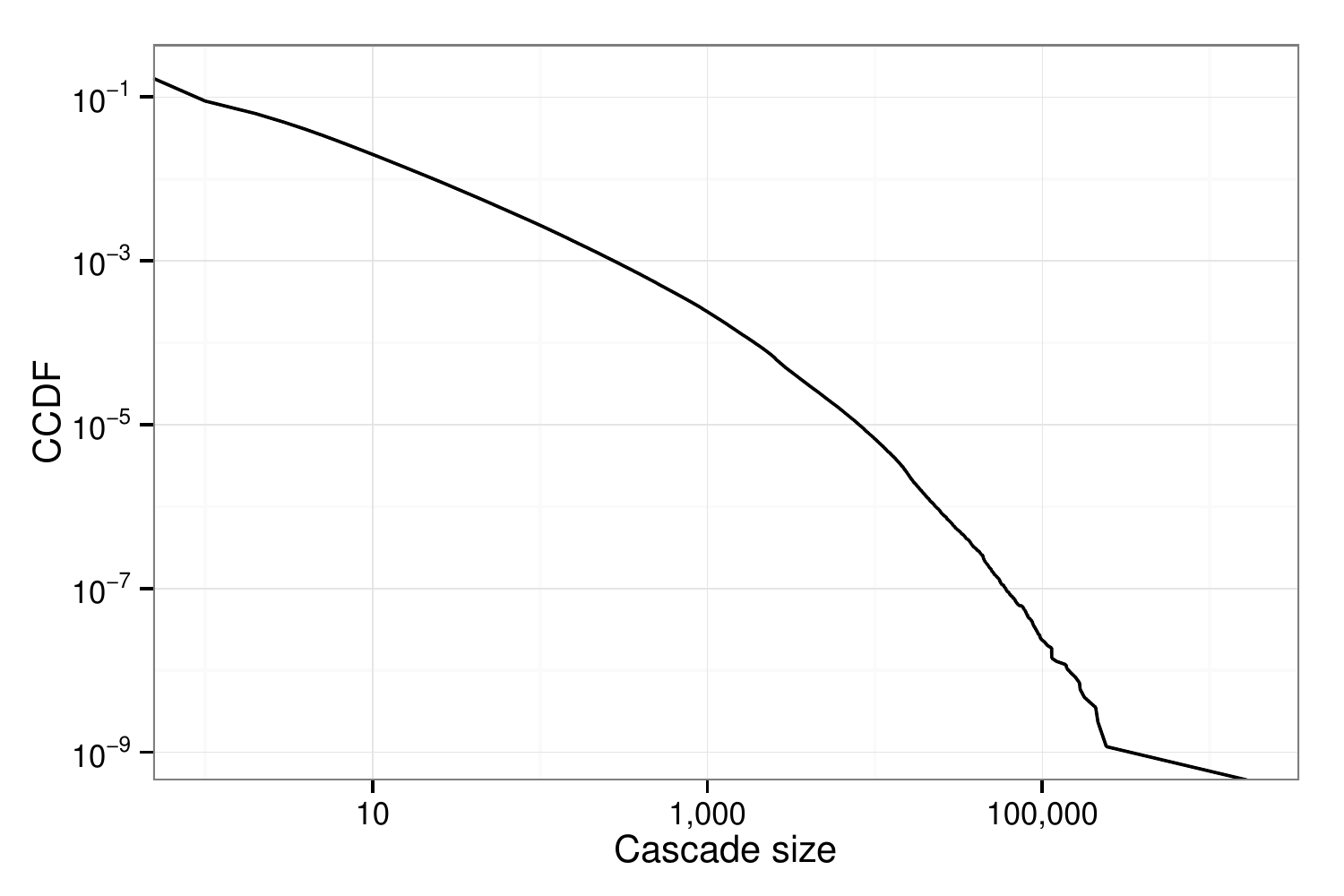}}
\caption{Complementary cumulative distribution of the number of followers for a Twitter user and cascade size of URLs.}
\label{fig:empirical_summary}
\end{figure*}

\begin{figure*}[ht]
\center
\subfloat{\includegraphics[scale=0.5]{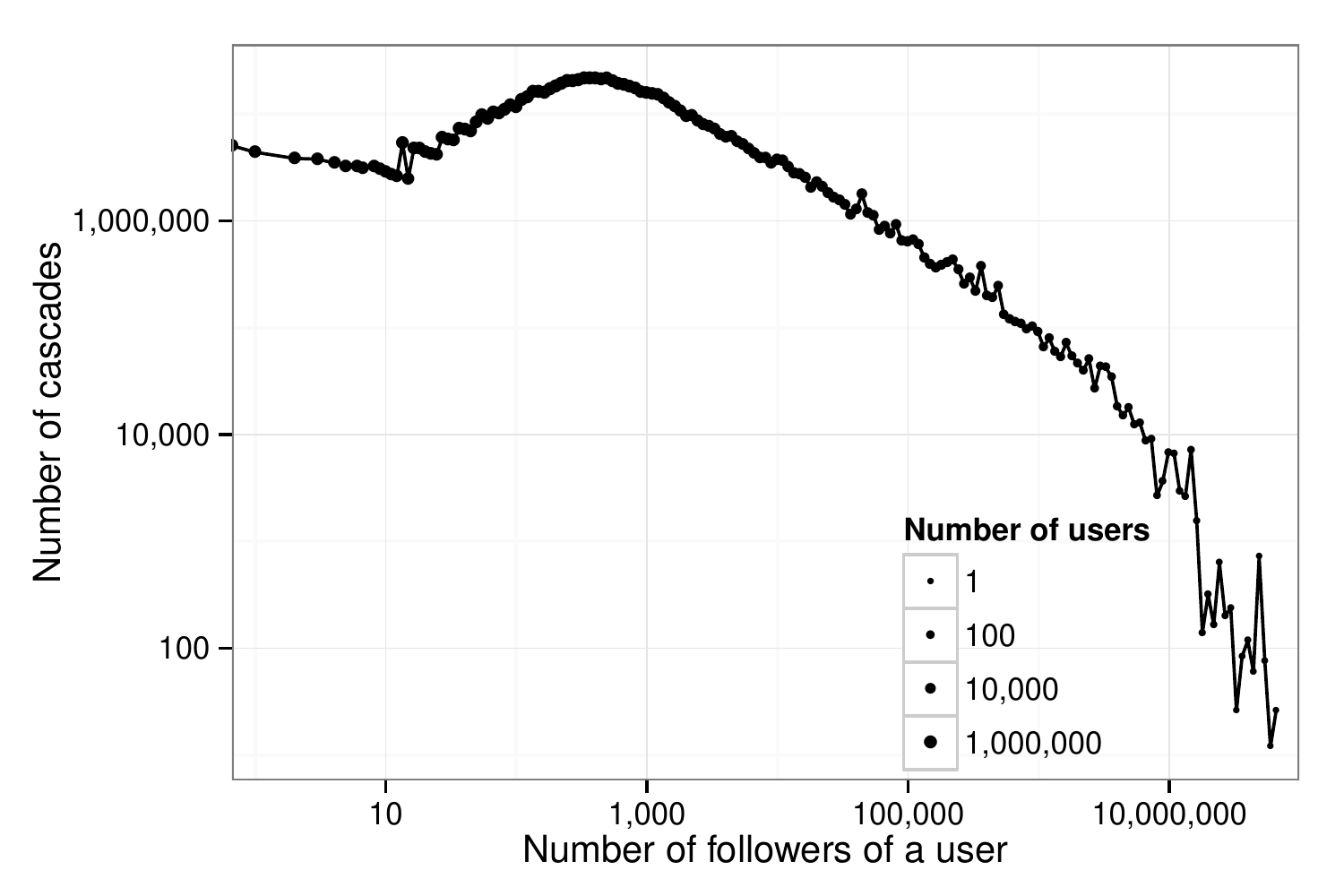}}
\quad
\subfloat{\includegraphics[scale=0.5]{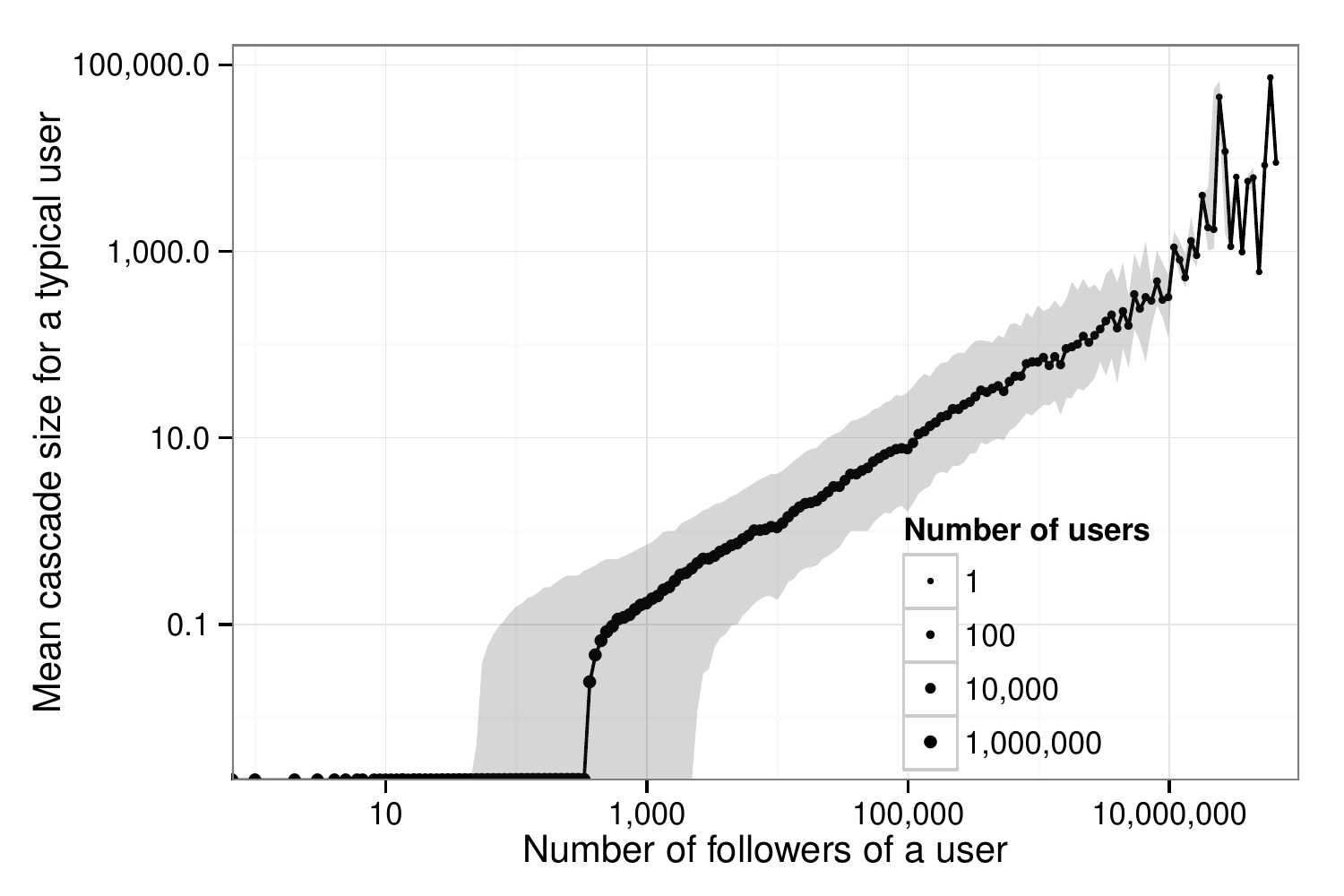}}
\caption{[Left panel] Average number of cascades seeded versus the degree of the seed user. [Right panel] Mean cascade size (retweets per tweet) for typical users at the median (line), 25th and 75th percentiles (lower and upper ribbon).}
\label{fig:empirical_summary_by_degree}
\end{figure*}

Figure~\ref{fig:empirical_summary} shows the distribution of follower count across all users in the data set (left) and the distribution of cascade sizes (right). In contrast to the general Twitter population, more than half of the users in our dataset have at least 100 followers, and the most followed user (currently Katy Perry) has 76 million followers. Cascades, on the other hand, typically involve a small number of users: fewer than 3\% of cascades are retweeted by 10 users or more.
Conditioning on the degree of seed users provides more insight on the distribution of cascades (Figure~\ref{fig:empirical_summary_by_degree}). Most cascades are started by users of low degree, owing to the higher frequency of low degree users in the dataset. On the other hand, the second panel of Figure~\ref{fig:empirical_summary_by_degree} shows that cascades started by higher degree users have larger reach on average. Thus, while users with around 100 followers generate a large fraction of the total cascades, an average tweet from one of these users is never retweeted. The skew in these distributions is typical of online activity and important to preserve in a realistic prediction task; random, balanced samples of popular and non-popular cascades may lead to convenient formulations \cite{petrovic2011,jenders2013,herremans2014} but will explain less of the observed variance in outcomes in complex systems like Twitter. Properly accounting for this skew will therefore be an important priority when we simulate cascades over social networks in Section~\ref{sec:simulation}.

\subsection{Features for prediction}
To approximate our notion of ``skill'' as accurately as possible we employ an extensive set of features, starting from basic content and user features and adding on advanced features that include topics and past successes for users and URLs.

\xhdr{Basic content features} We compute several simple features based on the content of each URL, including:
\begin{itemize}
		\adjustitemspace
		\item \domain, the domain of the URL contained in the tweet.
		\item \tweettime, the time at which a tweet was posted.
		\item \spamscore, as calculated by a proprietary Microsoft social media annotator.
		\item \tweetcategory, the OpenDirectoryProject\footnote{see \url{https://www.dmoz.org/}} subject area of the url, e.g. news, sports, or technology, as annotated by Microsoft.
\end{itemize}
\xhdr{Basic user features} We also use simple statistics about each user to capture their popularity and activity:
\begin{itemize}
		\adjustitemspace
		\item \followers, the tweeting user's follower count.
		\item \friends, how many users the user follows.
		\item \statuses, the user's current total tweet count.
		\item \usertime, the user's account creation time.
\end{itemize}

\xhdr{Topic features}
\label{sec:topics}
Although the above features have been used in past work to predict success~\cite{bakshy2011}, they are relatively coarse-grained and fail to capture more subtle details that differentiate users and content, as well as the interaction between them.
As such, we add more refined user and content topics to our feature set using a standard topic model~\cite{blei2003}.
Ideally we would do so by fitting a topic model to all of the content in our corpus, including tweet and URL text, from which we could compute topics for each tweet and each user (by aggregating over their tweets).
Given the size of our dataset, however, this approach is computationally prohibitive and we turn to a two-stage process to first assign users to topics and then categorize tweets.

We do so by taking advantage of data offered by Twitter's own users through Twitter Lists. That is, Twitter allows any user to create lists to bookmark a collection of other accounts. These lists are often thematic, and their titles have been shown to be effective indicators of the topics the users in the list tweets about~\cite{wu2011}. For example, users who tweet about data science tend to be on lists containing the words, ``data'' and ``science'', as well as related words such as, ``statistics'' and ``machine learning.'' As we describe next, this allows us to calculate the following features:
\begin{itemize}
			\adjustitemspace
			\item \usertopic, the topic about which a user generally tweets, 
			\item \tweettopic, the topic of any given tweet, and
			\item \interaction, the similarity between a given tweet and the user who introduced it.
\end{itemize}

\noindent \emph{User topics}.  Rate limit constraints make gathering all list data infeasible, so we assign topics to users by first using the Twitter API to collect list information for the most popular and active users on Twitter.
Specifically, we take the union of the top 10 million users sorted by follower count and the top 10 million users sorted by received retweets, yielding 16 million users who are popular, active, or both popular and active. 
We then crawl the first 1,000 lists that each of these users has been placed on, yielding a total of 23 million 
 unique lists and their titles.
Next, we fit a topic model to a user's list titles, as follows:
(1) we construct a ``document'' for each user by concatenating the titles of all of the lists they belong to; 
(2) after removing stop words and non-ASCII characters, we fit a Latent Dirichlet Allocation model using Mallet~\cite{mccallum2002} with 100 topics (to limit the computational requirements, we select a random subsample of 2 million non-empty documents for model fitting); and
(3) then use this trained model to infer topics for each of our 16 million users, and assign to each user their most probable topic.

\vspace{1.7mm} \noindent \emph{Tweet topics}. Next, we leverage the same 100 topics to categorize all tweets in our corpus according to the title text of each tweet's URL.
We do this by creating one document for each topic comprised of all of the URL titles for tweets posted by users belonging to that topic.
We treat this as a corpus of 100 documents and compute tf-idf scores to determine the relevance of all words to each topic.
Each tweet is then assigned to a topic based on the words in its corresponding URL title.
Specifically, each tweet is scored by the similarity of its content to the words in each topic, weighted by per-topic tf-idf scores, and the tweet is assigned to its highest scoring topic.

\vspace{1.7mm} \noindent \emph{Topic interaction.}
Finally, the user-tweet similarity score is defined to be 0 if user and tweet topics are different and the product of their topic weights otherwise. This can be interpreted as a \emph{topic interaction} term, reflecting the possibility that tweets that are ``on topic'' for that particular user will be received differently from those that are not (e.g. a user who is followed mostly for her expertise in data science may be unsuccessful when tweeting about a friend's art opening).

\xhdr{Past Success}
\label{sec:past_success}
For URLs that are tweeted multiple times, an alternative measure of potential to succeed (i.e. ``skill'') is the size of previous cascades.
Specifically, we compute past success of a URL by using a rolling average of the number of retweets received by the last 200 tweets containing that URL, using tweets up until the beginning of January. 
Correspondingly, a user's past success is defined in an analogous fashion, measuring a user's reach or her potential to start cascades of a certain size.  For each user, we compute the average number of retweets for the last 200 tweets posted by her.\footnote{As with the topic features, if a URL has not been tweeted in the past, or if user has no prior tweets, the value of this feature is null.} We thus obtain two features based on past success:
\begin{itemize}
		\adjustitemspace
		\item \urlsuccess, the average number of retweets received by tweets containing this URL in the past.
		\item \usersuccess, the average number of retweets received by this user in the past.	
\end{itemize}


\begin{table}
\begin{tabular} {cccc}
\textbf{Dataset} & \textbf{Users} &\textbf{ Tweets} &\textbf{ Retweets }\\
All tweets & 51.6M & 852M & 1.806B \\
Restricted tweets & 7.2M & 183M & 1.299B
\end{tabular}
\caption{Datasets used in empirical prediction exercise. Restricted tweets are those for which all user and content features are available.}
\label{tab:dataset_restriction}
\end{table}

\begin{table}[t]
\centering
\begin{tabular} {l|cccccccccccccc}
  \multicolumn{1}{c}{Model} & \rot{\tweettime} & \rot{\domain} & \rot{\spamscore} & \rot{\tweetcategory} & \rot{\tweettopic} & \rot{\urlsuccess} & \rot{\usertime} & \rot{\followers} & \rot{\friends} & \rot{\statuses} & \rot{\usertopic} & \rot{\usersuccess} & \rot{\interaction} \\ \hline
1.\ Basic content & \chk & \chk & \chk & \chk &&&&&&&&&\\
2.\ Content, topic & \chk & \chk & \chk & \chk & \chk&&&&&&&&\\
3.\ Content, past succ.\ & \chk & \chk & \chk & \chk & \chk & \chk &&&&&&&\\
4.\ Basic user &&&&&&& \chk & \chk & \chk & \chk & & & \\
5.\ User, topic &&&&&&& \chk & \chk & \chk & \chk & \chk & & \\
6.\ User, past succ.\  &&&&&&& \chk & \chk & \chk & \chk & \chk & \chk & \\
7.\ Content, user & \chk & \chk & \chk & \chk & \chk & \chk & \chk & \chk & \chk & \chk & \chk & \chk & \\
8. All & \chk & \chk & \chk & \chk & \chk & \chk & \chk & \chk & \chk & \chk & \chk & \chk & \chk \\
\end{tabular}

\caption{Features used in different models for cascade size prediction.}
\label{tab:features}
\end{table}

\begin{figure}[t]
\includegraphics[width=\linewidth]{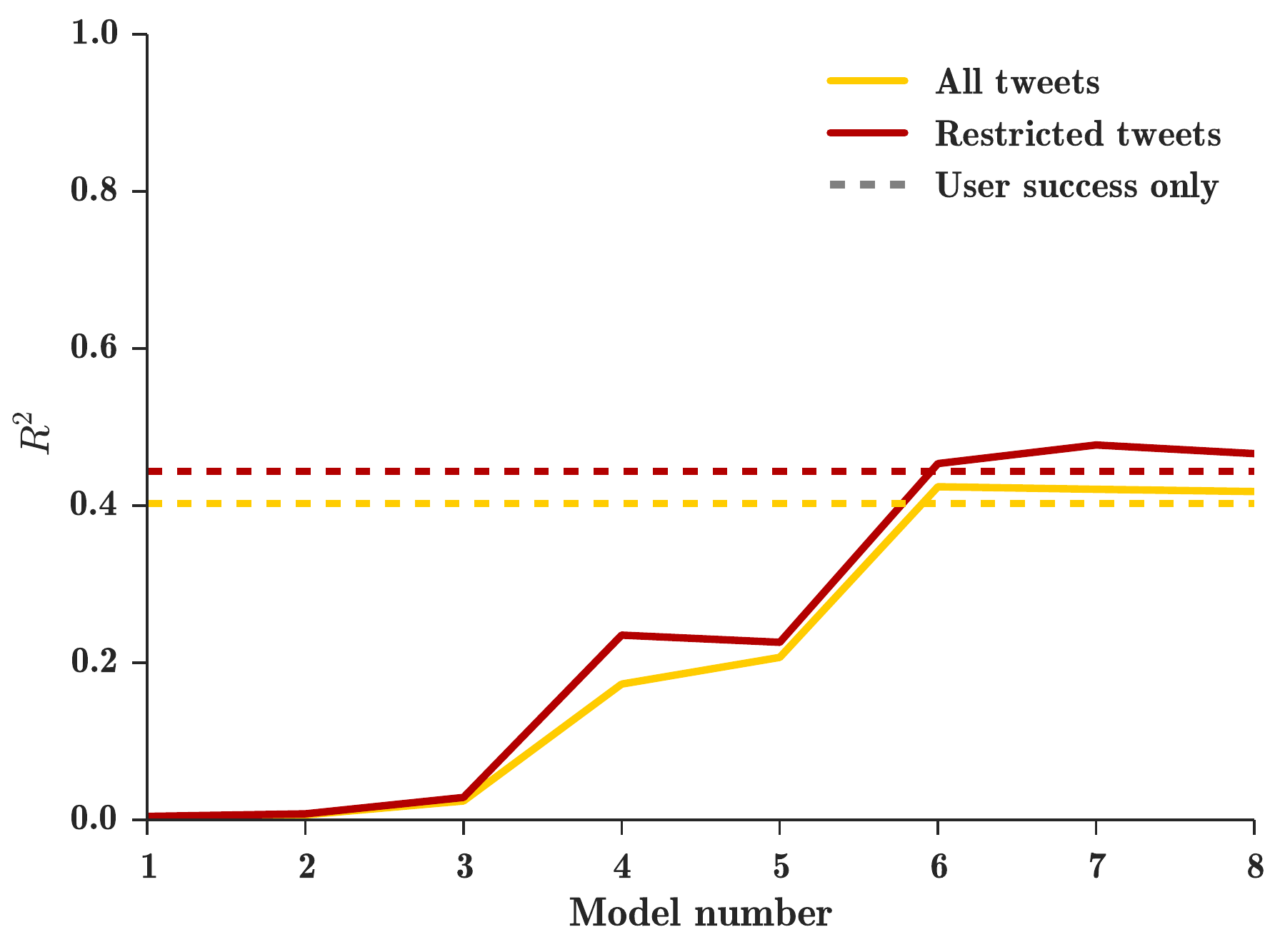}
\caption{Prediction results for models using different subsets of features. $R^2$ increases as we add more features, but only up to a limit. Even a model with all features explains less than half of the variance in cascade sizes.  }
\label{fig:empirical_perf}
\end{figure}

\subsection{Model evaluation}
We now turn to evaluating how well we can predict the size of URL cascades. We split our data chronologically into a training and test set, with the first three weeks of tweets (until 
February 20th) in the training set and the last week in the test set. We used a variety of statistical models to fit the data, including linear regression, regression trees and random forests. Random Forests (RF) consistently provided the most accurate predictions, thus we report results for the RF model through the rest of the paper. The RF model contains 10 trees per forest, a sampling rate of 0.8 and a tree depth of 11. As a measure against overfitting,  we stipulate that leaf nodes must contain at least 100 training examples.

We use two datasets for prediction, as shown in Table~\ref{tab:dataset_restriction}. The first one includes all 852 million tweets, while the second one is a subset of these tweets in which all user and content features are observable. Although this restriction greatly decreases the number of tweets and users in our dataset, Table~\ref{tab:dataset_restriction} shows that the remaining tweets still account for more than two-thirds of retweets, presumably because tweets introduced by the most active and highly followed users (i.e., those for whom we have the most features) attract more retweets than average.
To isolate effects of different features, we train the RF model using several subsets of tweet features, as described in Table~\ref{tab:features}.
We show prediction results for our models in Figure~\ref{fig:empirical_perf}, using $R^2$ to compare predicted and actual cascade size in the test dataset.    

The first set of models we evaluate are based on content features of a tweet. We find that content-based features alone perform poorly, consistent with previous work \cite{bakshy2011}: both basic content and topic features for a tweet lead to a negligibly low $R^2$. Even after including past success of tweets, $R^2$ stays well below $0.05$.
In comparison to content features of a tweet, features about the seed user  are more useful. For example, a model with only basic user features achieves $R^2$ close to 0.2, several times the performance of the best content-only model. Interestingly, including more advanced topic features for a user does not result in much improvement over and above the basic features. As with \cite{bakshy2011}, however, adding a user's past success leads to a large bump in model performance: $R^2$ increases to 0.42 on the unrestricted dataset, and up to 0.48 on the restricted dataset. Finally, adding more features over and above past user success (content features and user-tweet topic interaction) does not appreciably improve performance. This last result is somewhat surprising as intuitively one might expect that tweets that are ``on-topic''  for a particular user would perform better than off-topic ones; however, it may also be that past user success captures most of this effect. 

To summarize, performance increases as we consider more features, significantly outperforming previously reported \emph{ex-ante} prediction results~\cite{bakshy2011}, which achieved $R^2 \approx 0.34$. In other words, better estimates of ``skill'' combined with more powerful statistical models can improve predictive accuracy, as might be expected in a skill-based world. Balancing this optimistic result, we note that even with a much larger dataset and a far more extensive list of features (including e.g. user and content topics) than was available in~\cite{bakshy2011}, the best $R^2$ that can be achieved with a state-of-the-art model is 0.48, leaving more than half of the variance in cascade sizes unexplained. 
As with~\cite{bakshy2011}, moreover, a small number of user-based features account for most of this performance: in fact a single feature (past user success) performs almost as well on both datasets as all features combined.  

Clearly we cannot rule out that additional features might some day become available that will perform better than the features sets studied here, or that existing features can be combined in clever ways to boost performance, or that some other class of models might outperform random forests. 
Nevertheless, our finding that a relatively simple model, comprising just a single relatively coarse feature, can perform almost as well as more complex models suggests that much of the remaining unexplained variance is in fact not a consequence of an insufficiently rich model but rather derives from some intrinsic randomness in the underlying generative process. 
It is therefore this underlying generative process to which we now turn.

\section{Simulating Cascades}
\label{sec:simulation}
Complementing the empirical approach of the previous section, in this section we explore possible theoretical limits to \textit{ex-ante} prediction of success in a context similar to our empirical case. In practice, we simulate Twitter-like cascades in a ``model world'' in which we can directly control the contributions of skill and luck. In this way we can identify the best possible prediction algorithm---equivalent to our idealized model from Section~\ref{sec:model}---and thereby isolate the predictability of success given perfect \emph{ex-ante} knowledge of the system, as well as investigate how prediction performance degrades as one's knowledge becomes increasingly imperfect.

\subsection{Setting up the world}
At a high level, our model world comprises a network into which abstract ``products'' of varying ``appeal'' are introduced by some seed user and subsequently propagate to other users via some process of contagion, resulting eventually in a cascade of some size.
Given information about the initial seed, the network, and the appeal of a particular piece of content, 
the prediction task is to estimate the eventual size of each cascade.
We run simulations using a popular contagion model and use them to directly calculate asymptotic predictability limits as a function of knowledge of initial conditions.
To make these limits as informative as possible, we make a number of modeling choices, described below, to match our empirical data as closely as possible.


\xhdr{Network structure}  Our first objective is to approximate the Twitter follower graph (see e.g. Fig~\ref{fig:empirical_summary} and~\cite{bakshy2011,goel2015}) with as simple model as possible. 
To this end, we construct a static graph with 7 million nodes---much smaller than the actual Twitter graph, but similar in size to the network of active users we analyze---with a power-law distribution over node degrees with exponent $\alpha=2.05$, reflecting the best-fit to the empirical degree distribution (Figure~\ref{fig:empirical_summary}).

\xhdr{Contagion process} Next, we model contagion with a standard susceptible / infected / recovered (SIR) epidemic model that is widely studied in the diffusion of innovations literature~\cite{bass2004}, and is essentially the same as the independent cascade model~\cite{kempe2003} that is popular in the influence maximization literature. The SIR model proceeds in a sequence of rounds: in the first round, a single seed is infected with the contagion, and in subsequent rounds, the nodes infected in previous round infect their uninfected neighbors each independently with a fixed probability. The cascade stops spreading the first time there is a full round in which no new nodes are infected. As has been shown previously~\cite{goel2015} it is possible to adjust the parameters of an SIR model to match both the average size and full distribution of cascades sizes observed in Figure~\ref{fig:empirical_summary}.

\xhdr{Product quality (``skill'')}
As noted earlier, we interpret ``skill'' very broadly to mean some combination of intrinsic quality and contextual appeal, here associated with some ``product'' spreading through the network.
In our model world, we operationalize the quality/appeal of a particular product in terms of its \emph{reproduction number} $R_0$, defined as the expected number of new nodes infected by any individual node in a cascade, where in addition we assume that $R_0$ is sampled from a known probability distribution.
In light of previous work~\cite{goel2015} showing that Twitter cascades are best characterized by $R_0 \in [0,1]$, we sample reproduction numbers from a Beta distribution: $p(R_0) = \betadist(q, \sigma_q)$, where we have parameterized the distribution by its mean $q$ and standard deviation $\sigma_q$.

\xhdr{Seed users} Finally, when considering where to start cascades, we select a set of users that closely matches characteristics of seeds in our empirical Twitter data set.\footnote{We note that this distribution differs from the overall network's degree distribution: high-degree users are over-represented, seeding more cascades per capita than their low-degree counterparts.}
Specifically, we choose seeds to match the empirical distribution of activity by seed degree observed in \fref{fig:empirical_summary_by_degree}, as follows.
First, we use inverse transform sampling to sample the degree of a seed user.
Next, to adjust for the difference in network size between the empirical Twitter data and our simulated network, we match degrees by percentile between the real and simulated networks.
We then select a random node in the simulation graph with this percentile-adjusted degree.
In this way we can obtain a computationally feasible set of seed nodes sampled from a realistic activity distribution.
To cover a wide range of users and reproduction numbers, we simulate cascades for 10,000 different seed users and 800 uniformly spaced $R_0$ values. For each such seed user $u$ and reproduction number $R_0$, we simulate 1,000 independent cascades to obtain precise estimates of variance in cascade sizes.
This results in 8 billion simulations in total; we then post-stratify these cascades to match the relative frequencies in degree and $R_0$ distributions specified in the above section.

\begin{figure}[t]
\includegraphics[width=\linewidth]{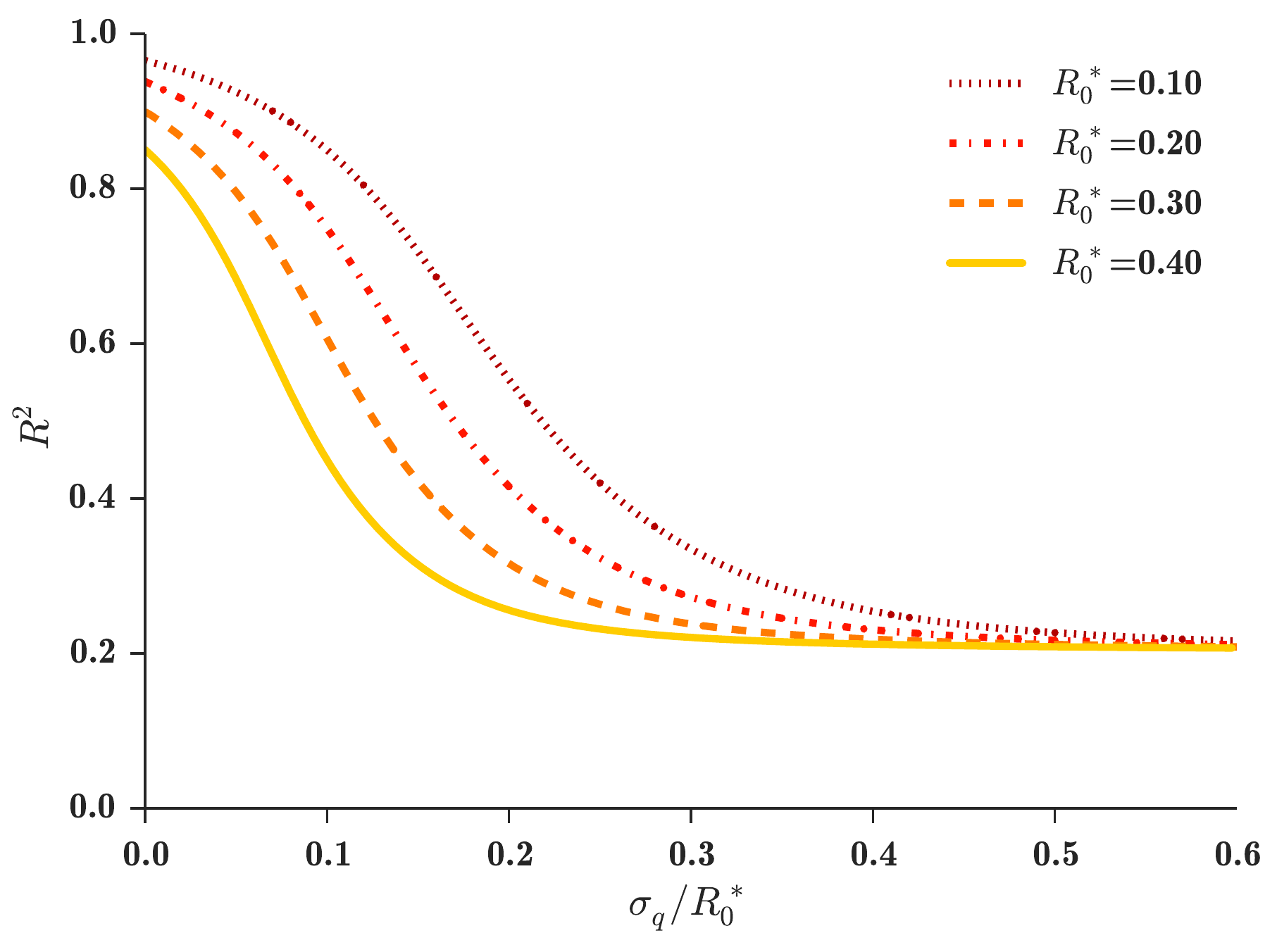}
\caption{Predictive performance in a simulated Twitter-like network assuming perfect knowledge about the world, for different distributions over product quality. As we introduce variation in product quality, predictability decreases. 
} 
\label{fig:sim_vary_beta}
\end{figure}

\begin{figure}[t]
\includegraphics[width=\linewidth]{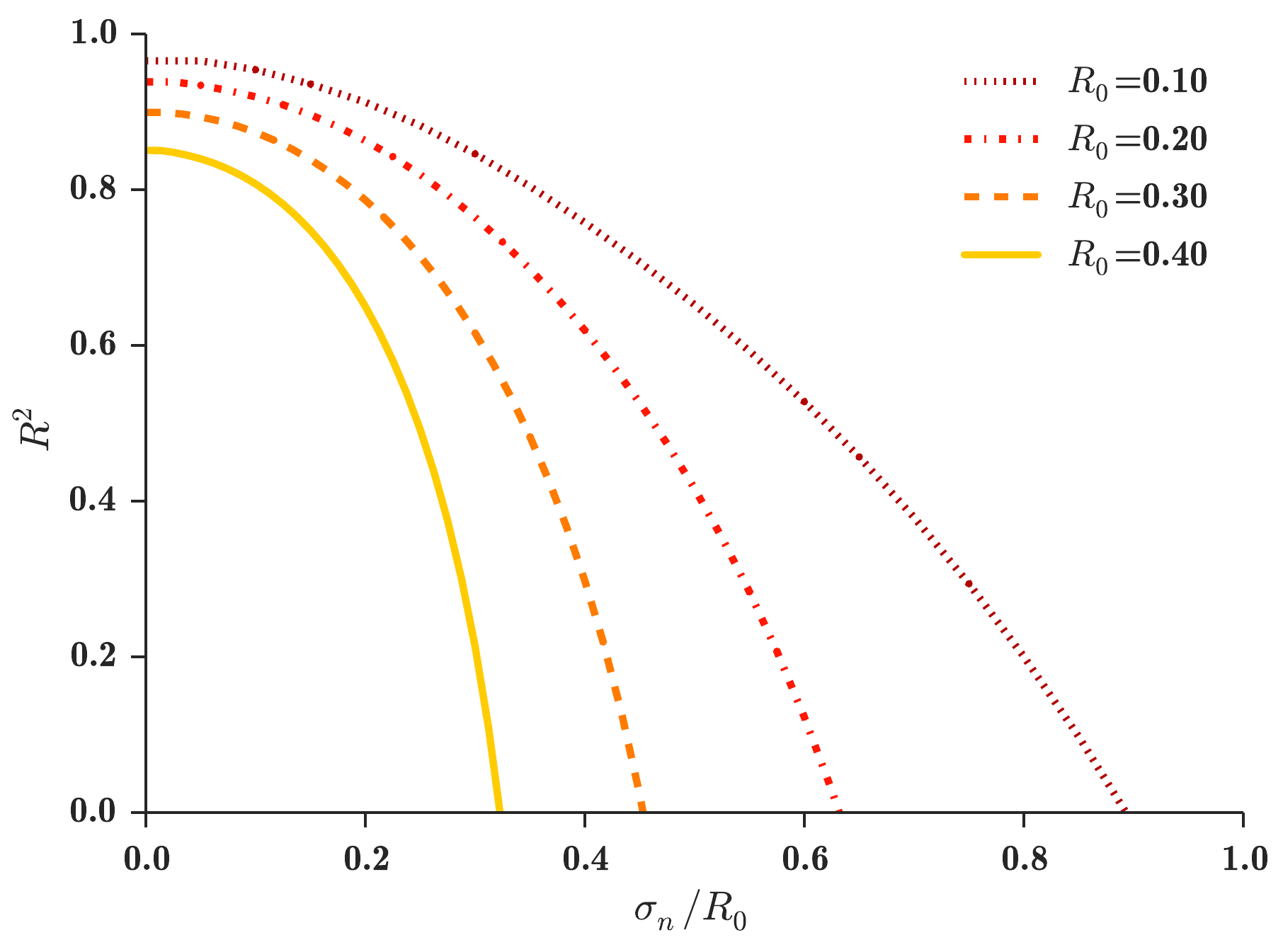}
\caption{Sensitivity analysis of how increasingly noisy estimates of quality affect prediction performance. As with variance in product quality, $R^2$ decreases with increase in the estimation error of $R_0$. }
\label{fig:sim_noise}
\end{figure}

\subsection{Limits on predictability of cascades}

Having specified our model world, we can investigate the variance in cascade sizes (success) given {\it identical} initial conditions, and thus put an upper bound on the predictability of success assuming perfect \emph{ex-ante} knowledge of the system.
Specifically we evaluate predictive performance in our simulations the same way we evaluated our empirical models: by measuring $R^2$, the amount of variance we can explain by conditioning on observable features.
In contrast to our empirical data, however, where quality is not directly observable, here we can simulate repeated cascades started by the {\it same user} with the {\it same infectiousness}, allowing us to condition on all sources of variation other than randomness of the diffusion process itself.
Specifically, we measure
\[ R^2 = 1 - \frac{\overline{\textrm{Var}}(S|R_0,u)}{\textrm{Var}(S)},
\]
where the numerator $\overline{\textrm{Var}}(S|R_0,u)$ is an estimate of the expected variance in success (cascade size) given knowledge of a cascade's propagation probability $R_0$ and seed $u$, and the denominator $\textrm{Var}(S)$ is the variance in success over all cascades.
The ratio between these two quantities describes the degree to which perfect \emph{ex-ante} knowledge of the seed user and the content quality $(R_0)$ conveys predictive power: if knowing the seed and $R_0$ is of negligible value, then $\textrm{Var}(S|R_0,u) \approx \textrm{Var}(S)$ and $R^2 \approx 0$; whereas if it explains almost all of the overall variance, then $\textrm{Var}(S|R_0,u) \ll \textrm{Var}(S)$ and $R^2 \approx 1$.


We first consider the scenario in which the quality of objects is known perfectly. The leftmost point on Figure~\ref{fig:sim_vary_beta} corresponds to the simplest case in which all products have the same quality (i.e. $\sigma_q=0$).  In such a restricted scenario, predictability is high, nearing $0.965$ for low $R_0$. At values of $R_0$ that are most appropriate for modeling the empirical distribution of cascade size on Twitter ($R_0 \in \{0.2,0.3\}$\footnote{We find that the mean cascade size in our empirical Twitter data is around 3, and this is reproduced in our simulations for $R_0$ between 0.2 and 0.3.}), the theoretical maximum prediction performance reduces to $0.93$.
Clearly, this limit seems high relative to what we can achieve in practice; however, we note that the dual assumptions of perfect knowledge and identical products is unlikely in the extreme. Nevertheless, it is instructive that even in this highly idealized scenario, predictive power is still bounded away from $R^2=1$. 

As we move further right in Figure~\ref{fig:sim_vary_beta}, we plot the slightly more realistic scenario in which knowledge is still perfect but product quality is increasingly allowed to vary (according to the $\betadist(q, \sigma_q)$ distribution described above).
We find that predictive performance decreases sharply as product variability increases. For instance, when $R_0$ is 0.20 on average and varies by just 15\% across products (corresponding to $\sigma_q/R_0^* = 0.15$ in Figure~\ref{fig:sim_vary_beta}), $R^2$ decreases to just 0.60.
The likely reason for this decrease is that increasing $\sigma_q$ increases the likelihood that cascades propagate with higher $R_0$'s, which tend to have larger outcomes with more variance, and are hence more unpredictable.
We emphasize that this inability to predict success given only product heterogeneity implies a some amount of ineradicable error in real-world predictions. That is, by conditioning on the actual user and product quality, we have eliminated possible shortcomings of any imperfect set of predictors in the real world, including our own in the previous section, such as lack of data, an insufficiently sophisticated model, and so on. Thus, any remaining errors must arise from the inherent unpredictability of the cascade process itself.






Finally, and building on this last observation, we note that any practical prediction task will also be hampered by imperfect \textit{ex-ante} knowledge of product quality. To quantify the impact of this constraint, we conduct a type of sensitivity analysis to determine how predictive performance degrades as knowledge of $R_0$ becomes imperfect. Cascades still propagate according to $R_0$, but now we assume that our prediction model has access only to a noisy estimate $\hat R_0 \sim \pN(R_0,\sigma_n)$ of this parameter. By varying $\sigma_n$, we can determine how sensitive our theoretical maxima established in the previous simulations are to knowledge of the precise $R_0$. For this setting, we define prediction performance as follows:
\[ R^2 = 1 - \frac{\overline{\textrm{Var}}(S|\hat R_0,u)}{\textrm{Var}(S)}, \]
where we condition on the estimated (instead of actual) reproduction number, $\hat R_0$.

Figure~\ref{fig:sim_noise} shows how performance degrades as $\sigma_n$ increases for various settings of $R_0$.
Again there are several salient conclusions we can draw from these simulations. First and foremost, prediction performance degrades quite rapidly as one's knowledge becomes even moderately noisy. For example, for $R_0=0.3$ (a value that closely recovers the average empirical cascade size), $R^2$ drops from 0.8 to below 0.6 as the standard deviation of noise ($\sigma_n$) increases from 
20\% to 30\% of $R_0$. This result complements our first set of findings from Figure~\ref{fig:sim_vary_beta}: although $R^2$ is bounded away from 1 given perfect knowledge of the system, introducing even a small amount of error in one's estimate of product quality has a dramatic negative effect on prediction performance.
We note, moreover, that in Figure~\ref{fig:sim_noise} we have chosen $\sigma_q=0$ (i.e. all products have equal $R_0$), which we, as we have already established, maximizes predictability in the perfect information case. Introducing heterogeneity in product quality decreases $R^2$  even further.
Combining results from Figures~\ref{fig:sim_vary_beta} and~\ref{fig:sim_noise}, therefore, we conclude that predictive performance in practical settings is limited for at least two distinct reasons:  first, there is inherent unpredictability even given perfect knowledge of the world---this is a luck world---and this limit is exacerbated by intrinsic heterogeneity in product quality; and second, even small deviations from perfect knowledge substantially decrease predictive performance.


\section{Discussion}
\label{sec:discussion}
Together, our empirical and simulation results suggest that there are genuine limits to \emph{ex-ante} prediction of the success of information cascades.
In our empirical work we find that our best model can explain less than half of the variance in cascades sizes on Twitter.
Although it remains unclear from the empirical results alone exactly how much of our model's performance is limited by estimation error or insufficient data as opposed to model mis-specification, our finding that past success alone performs almost as well as almost all other features combined suggests that adding more features is unlikely to result in large improvements. Moreover, we note that the volume and fidelity of data available for Twitter is almost certainly greater than for most real-world settings, hence whatever empirical limit we encounter in this case is likely to apply to other settings as well.

Further, our simulation results show that even with perfect knowledge of initial conditions---namely the infectiousness of a product and the identity of the seed user---there is inherent variability in the cascade process that sets a theoretical upper bound on predictive performance.
This bound is relatively generous for worlds in which all products are the same, but it becomes increasingly restrictive as we consider more diverse worlds with products of varying quality.
Moreover, these bounds on predictive performance are also extremely sensitive to the deviations from perfect knowledge we are likely to encounter when modeling real-world systems: even a relatively small amount of error in estimating
a product's quality
leads to a rapid decrease in one's ability to predict its success.


We see a number of possible limitations to our analysis and avenues for future investigation.
First, although we have calibrated our contagion model to reproduce the high level properties of our Twitter dataset such as degree distribution and cascade size, it is nonetheless a dramatic simplification of reality. We also note that there exists an extensive literature of contagion models that admit, among other features, more complex network properties such as clustering or homophily as well as complex contagion~\cite{romero2011}, non-uniform seeding~\cite{weng2013}, and limited attention spans~\cite{hodas2012,state2015}. Clearly more sophisticated models of this sort may be more realistic than the one we have studied, and may also yield somewhat different quantitative bounds to prediction. Thus although we anticipate that our \emph{qualitative} results will prove robust to our specific modeling assumptions, the relationship between model complexity and best-case predictive performance remains an interesting open question. 
%

Second, we have looked at only one measure of predictive performance in our empirical and theoretical work, and the choice of evaluation criterion is necessarily linked to what we might mean by predictability.
We chose $R^2$ because of its direct connections to our stylized model and its portability across domains; however, this choice also introduces particular sensitivities to how errors are quantified, namely sensitivity to outliers.
Considering other metrics, such as RMSE or MAE, might yield additional insights to the predictability of information cascades and avoid the issue of over-interpreting theoretical limits on $R^2$.
Likewise, variations on the problem formulation---e.g., \emph{ex-ante} prediction of the top-\emph{k} most popular cascades---might motivative different evaluation criteria and correspondingly reveal different bounds.

Finally, we have addressed only one domain and only one notion of success---namely the size of retweet cascades on Twitter. In articulating and framing the question, however, we hope that future work will examine limits to prediction in other domains and with other notions of success in a way that allows for meaningful comparisons. In this respect the broader contribution of our work is to clarify both the language used to talk about prediction in complex socioeconomic systems, and also the  often unstated assumptions underpinning many associated claims, such as what task is being studied, how performance is quantified, and what data is being used. In doing so, we hope to advance the current state-of-the-art in prediction as well as set expectations about what can be achieved even in principle and therefore to what extent predictions can be relied upon as input to planning and policy decisions~\cite{watts2011}.

\bibliographystyle{abbrv}
\bibliography{limits}

\begin{thebibliography}{10}

\bibitem{arrow2008}
K.~J. Arrow, R.~Forsythe, M.~Gorham, R.~Hahn, R.~Hanson, J.~O. Ledyard,
  S.~Levmore, R.~Litan, P.~Milgrom, F.~D. Nelson, et~al.
\newblock The promise of prediction markets.
\newblock {\em Science}, 320:877--878, 2008.

\bibitem{asur2010}
S.~Asur, B.~Huberman, et~al.
\newblock Predicting the future with social media.
\newblock In {\em International Conference on Web Intelligence and Intelligent
  Agent Technology}, volume~1, pages 492--499. IEEE, 2010.

\bibitem{bakshy2011}
E.~Bakshy, J.~M. Hofman, W.~A. Mason, and D.~J. Watts.
\newblock Everyone's an influencer: quantifying influence on {T}witter.
\newblock In {\em Fourth ACM international conference on Web search and data
  mining}, pages 65--74. ACM, 2011.

\bibitem{bass2004}
F.~M. Bass.
\newblock Comments on ``a new product growth for model consumer durables the
  bass model''.
\newblock {\em Management science}, 50(12):1833--1840, 2004.

\bibitem{bauer2015}
P.~Bauer, A.~Thorpe, and G.~Brunet.
\newblock The quiet revolution of numerical weather prediction.
\newblock {\em Nature}, 525(7567):47--55, 2015.

\bibitem{berger2013}
J.~Berger.
\newblock {\em Contagious: Why things catch on}.
\newblock Simon and Schuster, 2013.

\bibitem{berns2012}
G.~S. Berns and S.~E. Moore.
\newblock A neural predictor of cultural popularity.
\newblock {\em Journal of Consumer Psychology}, 22:154--160, 2012.

\bibitem{blei2003}
D.~M. Blei, A.~Y. Ng, and M.~I. Jordan.
\newblock Latent dirichlet allocation.
\newblock {\em The Journal of Machine Learning Research}, 3:993--1022, 2003.

\bibitem{bollen2011}
J.~Bollen, H.~Mao, and X.~Zeng.
\newblock Twitter mood predicts the stock market.
\newblock {\em Journal of Computational Science}, 2(1):1--8, 2011.

\bibitem{cheng2014}
J.~Cheng, L.~Adamic, P.~A. Dow, J.~M. Kleinberg, and J.~Leskovec.
\newblock Can cascades be predicted?
\newblock In {\em 23rd international conference on World wide web}, pages
  925--936. ACM, 2014.

\bibitem{choi2012}
H.~Choi and H.~Varian.
\newblock Predicting the present with {G}oogle trends.
\newblock {\em Economic Record}, 88(s1):2--9, 2012.

\bibitem{colizza2007}
V.~Colizza, A.~Barrat, M.~Barthelemy, A.-J. Valleron, A.~Vespignani, et~al.
\newblock Modeling the worldwide spread of pandemic influenza: baseline case
  and containment interventions.
\newblock {\em PLoS medicine}, 4(1):95, 2007.

\bibitem{demesquita2010}
B.~B. De~Mesquita.
\newblock {\em The Predictioneer's Game: Using the logic of brazen
  self-interest to see and shape the future}.
\newblock Random House Incorporated, 2010.

\bibitem{devany2004}
A.~De~Vany.
\newblock {\em Hollywood economics: How extreme uncertainty shapes the film
  industry}.
\newblock Routledge, 2004.

\bibitem{delsole2004}
T.~DelSole.
\newblock Predictability and information theory. part i: Measures of
  predictability.
\newblock {\em Journal of the atmospheric sciences}, 61(20):2425, 2004.

\bibitem{price1965}
D.~DeSolla~Price.
\newblock Networks of scientific papers.
\newblock {\em Science}, 149(3683):510--515, 1965.

\bibitem{domingos2015}
P.~Domingos.
\newblock {\em The Master Algorithm: How the Quest for the Ultimate Learning
  Machine will Remake our World}.
\newblock Basic Books, 2015.

\bibitem{frank2010}
R.~H. Frank and P.~J. Cook.
\newblock {\em The winner-take-all society: Why the few at the top get so much
  more than the rest of us}.
\newblock Random House, 2010.

\bibitem{friedman2010}
G.~Friedman.
\newblock {\em The next 100 years: a forecast for the 21st century}.
\newblock Anchor, 2010.

\bibitem{gardner2010}
D.~Gardner.
\newblock {\em Future Babble: Why Expert Predictions Fail---and Why We Believe
  Them Anyway}.
\newblock McClelland \& Stewart Limited, 2010.

\bibitem{ginsberg2009}
J.~Ginsberg, M.~H. Mohebbi, R.~S. Patel, L.~Brammer, M.~S. Smolinski, and
  L.~Brilliant.
\newblock Detecting influenza epidemics using search engine query data.
\newblock {\em Nature}, 457(7232):1012--1014, 2009.

\bibitem{goel2015}
S.~Goel, A.~Anderson, J.~Hofman, and D.~Watts.
\newblock The structural virality of online diffusion.
\newblock {\em Management Science}, 2015.

\bibitem{goel2010a}
S.~Goel, J.~M. Hofman, S.~Lahaie, D.~M. Pennock, and D.~J. Watts.
\newblock Predicting consumer behavior with web search.
\newblock {\em Proceedings of the National Academy of Sciences},
  107(41):17486--17490, 2010.

\bibitem{goel2010b}
S.~Goel, D.~M. Reeves, D.~J. Watts, and D.~M. Pennock.
\newblock Prediction without markets.
\newblock In {\em 11th ACM conference on Electronic commerce}, pages 357--366.
  ACM, 2010.

\bibitem{herremans2014}
D.~Herremans, D.~Martens, and K.~S{\"o}rensen.
\newblock Dance hit song prediction.
\newblock {\em Journal of New Music Research}, 43(3):291--302, 2014.

\bibitem{hodas2012}
N.~O. Hodas and K.~Lerman.
\newblock How visibility and divided attention constrain social contagion.
\newblock In {\em Conference on Social Computing (SocialCom)}, pages 249--257.
  IEEE, 2012.

\bibitem{holme2015}
P.~Holme and T.~Takaguchi.
\newblock Time evolution of predictability of epidemics on networks.
\newblock {\em Physical Review E}, 91(4):042811, 2015.

\bibitem{hong2010}
L.~Hong and B.~D. Davison.
\newblock Empirical study of topic modeling in {T}witter.
\newblock In {\em First Workshop on Social Media Analytics}, pages 80--88. ACM,
  2010.

\bibitem{hufnagel2004}
L.~Hufnagel, D.~Brockmann, and T.~Geisel.
\newblock Forecast and control of epidemics in a globalized world.
\newblock {\em Proceedings of the National Academy of Sciences},
  101(42):15124--15129, 2004.

\bibitem{ijiri1977}
Y.~Ijiri and H.~A. Simon.
\newblock {\em Skew distributions and the sizes of business firms}, volume~24.
\newblock North Holland, 1977.

\bibitem{jamali2009}
S.~Jamali and H.~Rangwala.
\newblock Digging digg: Comment mining, popularity prediction, and social
  network analysis.
\newblock In {\em International Conference on Web Information Systems and
  Mining}, pages 32--38. IEEE, 2009.

\bibitem{jenders2013}
M.~Jenders, G.~Kasneci, and F.~Naumann.
\newblock Analyzing and predicting viral tweets.
\newblock In {\em 22nd international conference on World Wide Web}, pages
  657--664. ACM, 2013.

\bibitem{kempe2003}
D.~Kempe, J.~Kleinberg, and {\'E}.~Tardos.
\newblock Maximizing the spread of influence through a social network.
\newblock In {\em Proceedings of the ninth ACM SIGKDD international conference
  on Knowledge discovery and data mining}, pages 137--146. ACM, 2003.

\bibitem{lazer2014}
D.~Lazer, R.~Kennedy, G.~King, and A.~Vespignani.
\newblock The parable of {G}oogle flu: traps in big data analysis.
\newblock {\em Science}, 343:1203--1205, 2014.

\bibitem{lerman2010}
K.~Lerman and T.~Hogg.
\newblock Using a model of social dynamics to predict popularity of news.
\newblock In {\em 19th international conference on World wide web}, pages
  621--630. ACM, 2010.

\bibitem{maity2015}
S.~K. Maity, A.~Gupta, P.~Goyal, and A.~Mukherjee.
\newblock A stratified learning approach for predicting the popularity of
  {T}witter idioms.
\newblock In {\em Ninth International AAAI Conference on Web and Social Media},
  2015.

\bibitem{mauboussin2012}
M.~J. Mauboussin.
\newblock {\em The success equation: Untangling skill and luck in business,
  sports, and investing}.
\newblock Harvard Business Press, 2012.

\bibitem{mccallum2002}
A.~K. McCallum.
\newblock Mallet: A machine learning for language toolkit.
\newblock http://mallet.cs.umass.edu, 2002.

\bibitem{orrell2008}
D.~Orrell.
\newblock {\em The future of everything: The science of prediction}.
\newblock Basic Books, 2008.

\bibitem{parish2006}
J.~R. Parish.
\newblock {\em Fiasco: A History of Hollywood's Iconic Flops}.
\newblock Wiley, 2006.

\bibitem{petrovic2011}
S.~Petrovic, M.~Osborne, and V.~Lavrenko.
\newblock {RT} to win! predicting message propagation in twitter.
\newblock In {\em ICWSM}, 2011.

\bibitem{pinto2013}
H.~Pinto, J.~M. Almeida, and M.~A. Gon{\c{c}}alves.
\newblock Using early view patterns to predict the popularity of {Y}outube
  videos.
\newblock In {\em Sixth ACM international conference on Web search and data
  mining}, pages 365--374. ACM, 2013.

\bibitem{polgreen2008}
P.~M. Polgreen, Y.~Chen, D.~M. Pennock, F.~D. Nelson, and R.~A. Weinstein.
\newblock Using internet searches for influenza surveillance.
\newblock {\em Clinical infectious diseases}, 47(11):1443--1448, 2008.

\bibitem{romero2011}
D.~M. Romero, B.~Meeder, and J.~Kleinberg.
\newblock Differences in the mechanics of information diffusion across topics:
  idioms, political hashtags, and complex contagion on {T}witter.
\newblock In {\em 20th international conference on World wide web}, pages
  695--704. ACM, 2011.

\bibitem{romero2013}
D.~M. Romero, C.~Tan, and J.~Ugander.
\newblock On the interplay between social and topical structure.
\newblock In {\em Seventh International AAAI Conference on Web and Social
  Media}, 2013.

\bibitem{salganik2006}
M.~J. Salganik, P.~S. Dodds, and D.~J. Watts.
\newblock Experimental study of inequality and unpredictability in an
  artificial cultural market.
\newblock {\em Science}, 311(5762):854--856, 2006.

\bibitem{schnaars1989}
S.~P. Schnaars.
\newblock {\em Megamistakes}.
\newblock Free Press; Collier Macmillan, 1989.

\bibitem{sherden1998}
W.~A. Sherden.
\newblock {\em The fortune sellers: The big business of buying and selling
  predictions}.
\newblock John Wiley \& Sons, 1998.

\bibitem{shulman2016}
B.~Shulman, A.~Sharma, and D.~Cosley.
\newblock Predictability of item popularity: Gaps between prediction and
  understanding.
\newblock Unpublished.

\bibitem{simonoff2000}
J.~S. Simonoff and I.~R. Sparrow.
\newblock Predicting movie grosses: Winners and losers, blockbusters and
  sleepers.
\newblock {\em Chance}, 13(3):15--24, 2000.

\bibitem{state2015}
B.~State and L.~Adamic.
\newblock The diffusion of support in an online social movement: Evidence from
  the adoption of equal-sign profile pictures.
\newblock In {\em 18th ACM Conference on Computer Supported Cooperative Work},
  CSCW '15, pages 1741--1750, New York, NY, USA, 2015. ACM.

\bibitem{surowiecki2005}
J.~Surowiecki.
\newblock {\em The wisdom of crowds}.
\newblock Anchor, 2005.

\bibitem{szabo2010}
G.~Szabo and B.~A. Huberman.
\newblock Predicting the popularity of online content.
\newblock {\em Communications of the ACM}, 53(8):80--88, 2010.

\bibitem{taleb2010}
N.~N. Taleb.
\newblock {\em The black swan: The impact of the highly improbable}.
\newblock Random House, 2010.

\bibitem{tetlock2005}
P.~Tetlock.
\newblock {\em Expert political judgment: How good is it? How can we know?}
\newblock Princeton University Press, 2005.

\bibitem{tetlock2015}
P.~E. Tetlock and D.~Gardner.
\newblock {\em Superforecasting: The art and science of prediction}.
\newblock Crown, 2015.

\bibitem{watts2011}
D.~J. Watts.
\newblock {\em Everything is obvious:* Once you know the answer}.
\newblock Crown Business, 2011.

\bibitem{weaver1961}
W.~Weaver.
\newblock A quarter century in the natural sciences.
\newblock {\em Public health reports}, 76(1):57, 1961.

\bibitem{weng2013}
L.~Weng, F.~Menczer, and Y.-Y. Ahn.
\newblock Virality prediction and community structure in social networks.
\newblock {\em Scientific reports}, 3, 2013.

\bibitem{weng2014}
L.~Weng, F.~Menczer, and Y.-Y. Ahn.
\newblock Predicting successful memes using network and community structure.
\newblock In {\em Eighth International AAAI Conference on Weblogs and Social
  Media}, 2014.

\bibitem{wu2011}
S.~Wu, J.~M. Hofman, W.~A. Mason, and D.~J. Watts.
\newblock Who says what to whom on {T}witter.
\newblock In {\em 20th International Conference on World Wide Web}, pages
  705--714. ACM, 2011.

\bibitem{yu2015}
L.~Yu, P.~Cui, F.~Wang, C.~Song, and S.~Yang.
\newblock From micro to macro: Uncovering and predicting information cascading
  process with behavioral dynamics.
\newblock {\em IEEE International Conference on Data Mining}, 2015.

\bibitem{zhao2015}
Q.~Zhao, M.~A. Erdogdu, H.~Y. He, A.~Rajaraman, and J.~Leskovec.
\newblock Seismic: A self-exciting point process model for predicting tweet
  popularity.
\newblock In {\em 21th ACM SIGKDD International Conference on Knowledge
  Discovery and Data Mining}, pages 1513--1522. ACM, 2015.

\end{thebibliography}

\end{document}